\documentclass[a4paper,aps,twocolumn,floats,showpacs,superscriptaddress]{revtex4}

\usepackage{epsfig}

\begin{document}

\title{Time evolution of damage under variable ranges of load transfer}

\author{Oluwole E. Yewande} 

\affiliation{The Abdus Salam International
Centre for Theoretical Physics, P.O. Box 586, Trieste, I-34014, Italy}

\author{Yamir Moreno} 

\altaffiliation{Address after November 2002: Departamento de
F\'{\i}sica Te\'orica, Universidad de Zaragoza, Zaragoza 50009, Spain}

\affiliation{The Abdus Salam International Centre for Theoretical
Physics, P.O. Box 586, Trieste, I-34014, Italy}

\author{Ferenc Kun}

\affiliation{Department of Theoretical Physics, University of
Debrecen, P.O.Box: 5, H-4010 Debrecen, Hungary}

\author{Raul Cruz Hidalgo}

\affiliation{Institute for Computational Physics, University of
Stuttgart, Pfaffenwaldring 27, 70569 Stuttgart, Germany}

\author{Hans J. Herrmann} 

\affiliation{Institute for Computational Physics, University of
Stuttgart, Pfaffenwaldring 27, 70569 Stuttgart, Germany}

\date{\today}


\begin{abstract} 
We study the time evolution of damage in a fiber bundle model in which
the range of interaction of fibers varies through an adjustable stress
transfer function recently introduced. We find that the lifetime of
the material exhibits a crossover from mean field to short range
behavior as in the static case. Numerical calculations showed that the
value at which the transition takes place depends on the system's
disorder. Finally, we have performed a microscopic analysis of the
failure process. Our results confirm that the growth dynamics of the
largest crack is radically different in the two limiting regimes of load
transfer during the first stages of breaking.
\end{abstract}



\maketitle

\section{Introduction}

The phenomenon of fracture in heterogeneous materials is a complex
physical problem which has been of great interest to scientists for
quite a long time \cite{r1,r2,r3}. A disordered system is understood
to be one with random time or/and space dependent breaking properties
\cite{r1}. The random nature of this dependence arises because modeling
the fracture of heterogeneous materials entails dealing with systems
made up of many constituents, each one having mechanical properties
that can be considered as being independent of the properties of the
other constituents, but with many body interactions among the
different parts of the system \cite{r1,r2,r3}. In heterogeneous
materials the evolution of the rupture process is radically different
from the single crack growth mechanism that occurs in homogeneous
materials. Though there is neither a complete numerical solution nor
analytic solution to the fracture problem, we have a better
understanding of it due to some recent algorithms that have been
developed to simulate the fracture process \cite{r4,r5,r6,r7,r8}.

Fiber Bundle Models (FBM) form a fundamental class of approaches to
the fracture problem. They arose in close connection with Daniels' and
Coleman's seminal works on the strength of bundles of textile fibers
\cite{r9,r10}, and have harbored an intense research activity in
recent years
\cite{r1,r2,r3,r4,r11,r12,r13,r14,r15,r16,r17,r18,r19,r19b,r19c}. Fiber
Bundle Models are important, despite their very simple nature, because
they exhibit most of the essential aspects of material breakdown. In
addition, the deep understanding of fracture processes they provide
has served as a starting point for more complex models
\cite{r13,r14,r15,r19a,r20,r21}. In FBMs the material is made up of a
set of parallel fibers, each having a statistically distributed
threshold strength or life-time. Besides the classical static FBM one
can also introduce a dynamic version \cite{r18}. In the static FBM,
the failure process is simulated according to a quasi static loading
of the material in which the output of the simulation is the ultimate
strength of the material, i.e, the maximum load above which it breaks
down. In the dynamic FBM, a constant load is maintained on the system
and the fibers break by fatigue after some time.

After a fiber breaking, the load acting on it is redistributed among
the intact fibers according to the stress redistribution rule
associated with the particular load-transfer model. There are two
standard load transfer models comprising the FBM and they correspond
to the extreme limits of stress redistribution. The global load
sharing (GLS) redistributes the load of a failed fiber equally among
the active fibers remaining in the system. It is known as the global
fiber bundle model (GFBM), and assumes long-range interaction among
the fibers which makes it a mean-field approximation that can be
solved analytically \cite{r11,r12}. On the other hand, the local load
sharing (LLS) redistributes the load of a failed fiber among the
intact fibers that are nearest neighbors to the failed ones, and is
thus known as the local fiber bundle model (LFBM). This assumes
short-range interaction among the fibers and it has not, in general,
been solved analytically. However, in actual heterogeneous materials
the stress redistribution is expected to fall in between the two
regimes of load transfer. Very recently, some of us have proposed a
load transfer scheme with variable range of interaction among the
fibers, which interpolates between the two extreme cases.  Since most
of the physics of the fracture problem is hidden in the stress
redistribution, considering within the same model the possibility of
varying the range of interaction is a relatively greater improvement
toward a better understanding of the fracture problem.

Motivated by the results obtained for the static setting, we have
studied a stochastic dynamic fiber bundle model, in which fibers break
by fatigue over a period of time, such that the range of interaction
among fibers is variable through an adjustable stress transfer
function. We have observed a crossover from mean-field to short-range
behavior for the macroscopic quantities describing the fracture
dynamics as in the static case. In addition, a more detailed
inspection of the failure process revealed that the microscopic
behavior of the system is also different as the range of interaction
varies. Finally, we have also studied the effect of heterogeneity on
the failure process from macroscopic as well as microscopic
perspectives. The rest of the paper continues with the description of
the stochastic model in the next section. Section\ \ref{stf} is
devoted to study the lifetime of the bundle and the rate at which
fibers break when the range of interaction changes. The damage
spreading for several load transfer modalities is addressed in
section\ \ref{scg} while the last section is devoted to conclusions
rounding off the paper.

\section{The stochastic model.}
\label{ssm}

We assign each fiber to the sites of a square lattice, with periodic
boundary conditions. Assuming elastic
interaction among the fibers, we may state that the stress
redistribution obeys a power law \cite{r21},
\begin{equation}
\sigma_{inc} {\sim} \frac{1}{r_{ij}^{\gamma}},
\label{eq1}
\end{equation} 
where $\sigma_{inc}$ is the load increment on an intact fiber $i$ at a
distance $r_{ij}$ from the failed fiber $j$, and $\gamma$ is a variable
parameter that controls the effective range of interaction among the
fibers. $\gamma$=0 corresponds to GLS since the additional load on
each intact fiber as a result of a fiber breaking is the same
irrespective of its distance from the broken fiber. On the other hand,
$\gamma=\infty$ corresponds to LLS since in this limit only the
nearest neighbors get equal portions of the load of a failed fiber,
and $\sigma_{inc}$=0 for $r>1$. Assuming that there is no
dissipation of the load of a failed fiber, Eq.\ (\ref{eq1}) leads us
to the normalized stress transfer function,
\begin{equation}
S(\gamma,r_{ij})=\frac{1}{r_{ij}^\gamma}
\frac{1}{\sum_{i\in{A}}r_{ij}^{-\gamma}}
\label{eq2}
\end{equation}
$r_{ij}$ being the distance between an active fiber $i$, with
coordinates $(x_i,y_i)$, and a failed fiber $j$, with
coordinates $(x_j,y_j)$, and $A$ denotes the set of intact fibers.

One can consider two equivalent approaches to the dynamic FBM
\cite{r22,r23,r24}. The time elapsed until the final collapse of the
system is the lifetime or time to failure of the bundle. In the first
approach \cite{r22} the lifetime of each element is an independent
identically distributed random variable, i.e, each fiber has a
different random lifetime, taken from the same statistical
distribution (the Weibull distribution is a good empirical
distribution in materials science) and each one breaks if the time
elapsed exceeds its individual lifetime. This is a {\em quenched}
model of fracture where the disorder is introduced once for all at the
beginning of the process and thus the growth mechanism is
deterministic. In this version, the effect of the increase in stress
for a particular fiber $i$ due to the redistribution of load from
failed fibers is the reduction of its initially assigned lifetime
$t_i$ to a new lifetime $\tau_i$ given by
\begin{equation}
t_i=\int_{0}^{\tau_i}\left(\frac{\sigma_i(t)}{\sigma_0}\right)^{\rho}dt,
\label{eq1a}
\end{equation}
where $\sigma_0$ is the initial stress on fiber $i$ at $t=0$ and
$\rho$ is the Weibull index, $2\le\rho\le50$. $\rho$ gives the degree
of heterogeneity of the system; as $\rho$ increases the system becomes
more homogeneous. In this model the next element to break is exactly
the one with the lowest lifetime at time $t$.

In the probabilistic approach \cite{r23,r24}, it is considered that in
a time step all the intact elements have the same mean time to
failure. The element that breaks in the time interval between two
successive failures is selected by chance and thus the fiber whose
probability of breaking (a function of the load it bears) is the
largest is more likely to fail. The probabilistic approach is an
example of the so called {\em annealed} disorder since the algorithm
is stochastic and thus randomness is uncorrelated in time. Thus, we
start at time $t=0$ with $N$ fibers loaded with an initial common
stress of $\sigma_i(t)=\sigma_i(0)=1$.  The mean time interval
$\delta$ for an individual element to break by fatigue is,
\begin{equation}
\delta=\frac{1}{\sum_{i\in{A}}\sigma_i^{\rho}(t)},
\label{delta}
\end{equation}
where $A$ is the set of intact fibers and $t$ is the time elapsed up to $k-1$
breakings, i.e., $t=\sum_{p=1}^{k-1}\delta_p$. In the first time interval
where $\sigma_i=1$ $\forall i$, with all the $N$ elements active,
$\delta_1=\frac{1}{N\sigma_i^\rho}$ so that the first fiber breaking is
completely random. This changes with time due to stress transfers and
fiber failures. The probability of a particular fiber breaking $j$ in
one sweep of the lattice in the time interval $\delta_k$, is given by
\begin{equation}
P_j(t)=\sigma_j^{\rho}(t)\delta_k .
\label{eq4}
\end{equation}

When a fiber breaks, its load is transfered and as a consequence, the
probability of failure of the receptors is increased. In this sense
there are no weak or strong fibers, all the fibers are equivalent but
carrying different loads (except for mean field approaches) and thus
with different breaking probabilities. It is worth stressing that both
approaches are equivalent. This can be intuitively understood by
noting that in both models the load history plays a key role. For the
{\em quenched} setting the fiber that breaks is that with the lowest
lifetime which in its turn depends on the load history. Since the
individual times to failure are reduced each time a fiber receives
load from a failed element, the more stressed a fiber is, the
more likely its lifetime is the lowest. On the other hand, for the
{\em annealed} version the magnitude that is modified by the load
redistribution is the probability of breaking and the load history
affects the failure process just as in the deterministic version: the
more stressed a fiber is, the more likely it breaks. Additionally, we
note that this algorithm is the same as the one used in polymer failure
\cite{pol} and in describing dielectric breakdown \cite{db} with the
main distinction that here the broken fibers need not to be connected
to the single growing cluster as in dielectric breakdown.

Now, the load born by the fiber that has just failed is redistributed
according to Eq.\ (\ref{eq1}) such that in a time interval $\Delta
t=t_k-t_{k-1}=\sum_{p=1}^k\delta_p-\sum_{p=1}^{k-1}\delta_p=\delta_k$,
the load increase on the active fiber $i$ reads as
\begin{equation}
\sigma_i(t_k)=\sigma_j(t_{k-1})S(\gamma,r_{ij})+\sigma_i(t_{k-1}),
\end{equation} 
where $\sigma_j$ is the load of
the element that has failed in the time interval ${\Delta}t$ after $k$
fiber failures. After
the redistribution of load the rupture process continues by applying
again Eq.\ (\ref{eq4}) that will point to the next fiber to break. The
lifetime of the material is then given by the sum of all the $N$
$\delta$s, $T_f=\sum_{i=1}^N\delta_i$.

The complete analytic solution to the dynamic FBM as defined above in
the GLS ($\gamma=0$) case is feasible. The lifetime of the material
can be computed as
\begin{equation}
T_f=\int_1^N\frac{(N+1-x)^{\rho-1}}{N^{\rho}}dx=\frac{1}{\rho}\left[1-\frac{1}{N^\rho}\right]
\label{eq6}
\end{equation} 
that includes also the dependence of the lifetime on the system size
$N$ giving the mean field result $1/\rho$ in the thermodynamic limit.

We should note that there is no avalanche here, unlike to the static
case \cite{r21}, since there is no external driving on the
system. Once the fibers start breaking by fatigue, they continue
breaking with time until the final collapse of the system, but within
a time interval $\delta_k$, only one fiber may break in a single sweep
of the lattice.
 
We note that both Eqs.\ (\ref{eq1a}) and (\ref{eq4}) assume a
power-law breaking rule
$\kappa_p(\sigma)=\nu_0(\sigma/\sigma_0)^{\rho}$. This stems from the
former assumption that the lifetimes satisfy the Weibull
distribution. An alternative breaking rule could be and exponential
hazard rate of the form
$\kappa_e(\sigma)=\phi\exp(\eta(\sigma/\sigma_0))$ mainly used for
thermally activated processes. On the other hand, $\kappa_p$ has the
same functional form as the Charles power law that describes the
subcritical crack growth induced by stress corrosion in geological
materials at constant temperature: $v=AK_I^n$, where $v$ is the crack
velocity and $K_I$ is the stress intensity factor for mode $I$
fracture. Sometimes, $n$ is known as the stress corrosion
index. Moreover, we emphasize that Eq. (\ref{delta}) is not a real
measure of time so that a quantitative comparison between the results
here shown and those from experiments is not feasible. Additionally,
Eq. (\ref{delta}) is a simple form one can consider for the
probability of breaking. In principle, one could also include more
realistic rules. Nevertheless, as we shall see, this
simple model is very rich and might help understand physical effects
present in real materials.

\section{Lifetime of the bundle} 
\label{stf}

We simulate the failure process by large scale numerical
simulations. We first explore the behavior of the lifetime as a
function of the stress transfer range for a fixed level of
heterogeneity ($\rho=2$). Then we vary the Weibull index such that we
get a more homogeneous system with smaller times to failure.
\begin{figure}[t]
\begin{center}
\epsfig{file=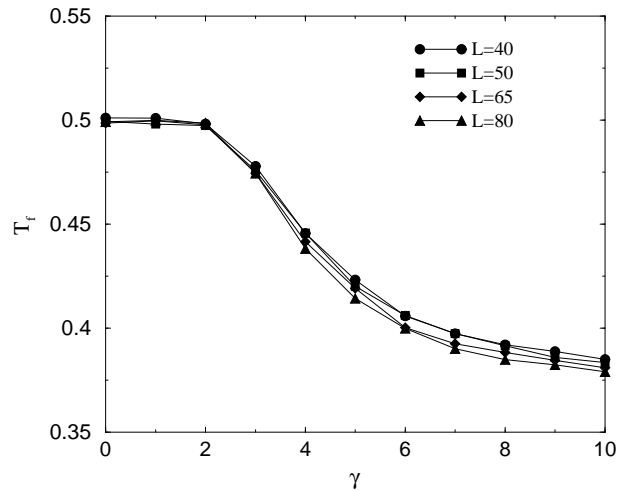,width=2.6in,angle=-90,clip=1}
\end{center}
\caption{Lifetime $T_f$ versus effective range of interaction
$\gamma$. The crossover from mean-field to short-range
behavior is obtained at the same transition point as in the static
case ($\rho=2$).}
\label{fig1}
\end{figure}
The results obtained for the lifetime are depicted in Fig.\ \ref{fig1}
for different values of the range of interaction among the fibers and
several system sizes up to $N=6400$ fibers. A crossover from mean
field behavior to a regime dominated by the short range interactions
among the fibers is clearly appreciated. Furthermore, the critical
value of the parameter $\gamma_c$ defines a region where the results
for global load sharing models hold beyond $\gamma=0$, the true
mean-field regime for which the load of a failed element is shared
equally among the surviving elements. For $\gamma \leq \gamma_c$ the
material behaves macroscopically as for $\gamma=0$, that is, the
lifetime is independent on the system size and does not depend on the
actual value of the exponent $\gamma$. It is not a simple numerical
task to determine accurately the exact value of the transition point
due to the stochastic nature of the model and the fluctuations of the
lifetime of the system. In fact, the time to failure of the bundle
follows a Gaussian distribution in what concerns its frequency
distribution. The width of the distribution depends on the level of
heterogeneity (controlled by the Weibull index) that in turn also
influences the lifetimes that become shorter as we move to high levels
of homogeneity. Within this numerical uncertainty we have found that
$\gamma_c \sim 2$. Interestingly, the same value was found to
characterize the transition of the bundle's ultimate strength from
long to short range behavior in the static case of the model
\cite{r21}.

\begin{figure}[t]
\begin{center}
\epsfig{file=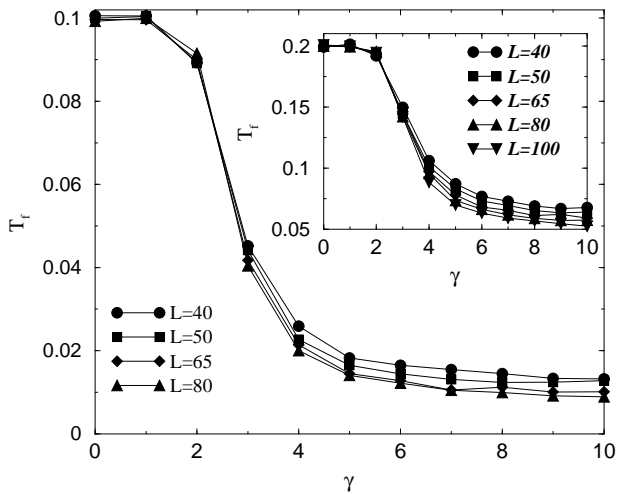,width=2.6in,angle=-90,clip=1}
\end{center}
\caption{Comparison of the time to failure obtained for different
heterogeneity levels $\rho$ as the range of interaction varies. The
inset corresponds to $\rho=5$ while the main figure has been drawn
considering $\rho=10$. Clearly, the crossover behavior is still
present although the value of the transition point depends on
$\rho$. As the system gets more homogeneous, $\gamma_c$ shifts
leftwards.}
\label{fig2}
\end{figure}

The influence of the disorder on the crossover behavior can be studied
by changing the value of $\rho$. Fig.\ \ref{fig2} shows the time to
failure of the material as a function of the effective range of
interaction for several system sizes. We observe that the transition
is still present but the range where the mean field regime applies is
reduced. In particular, as the system gets more homogeneous $\gamma_c$
shifts leftwards to smaller values. Additionally, the true local load
sharing regime appears to be also slightly shifted to the left. When
the range of interaction $\gamma$ falls below a second transition point
$\gamma_{c2}$, the lifetime of the bundle becomes again independent on
the effective interaction among the fibers but it is still size
dependent. This later behavior can be easily understood by noting that
for local load sharing schemes the time to failure of the system in
the thermodynamic limit is zero. We have checked that in our model
this is actually verified, although the drop of the lifetime as the
system size is increased is slow. On the other hand, wherever the GLS
regime arises, the time to failure of the system is not size dependent
for large $N$ (see Eq.\ (\ref{eq6}).)

Another way to characterize the evolution of the fracture process is
to inspect the rate at which fibers fail. We expect two different
asymptotic regimes. For long range interaction, {\em i.e.}, below
$\gamma_c$ the system should behave in a mean field manner. This
means that damage is gradually accumulated in the material up to a
point in which the load is too high as to be carried by the remaining
fibers. Only at this stage of the damage process the rate of fiber
failures will speed up owing to the small values of the very last
$\delta$s. On the contrary, in the region where short range
interaction prevails, the system does not accumulate damage
uniformly. In this case there appear regions within the material in
which stress enhancement takes place making the fibers along the crack
tips to support much more load than other active fibers placed far
from the clusters of broken fibers. 
\begin{figure}[t]
\begin{center}
\epsfig{file=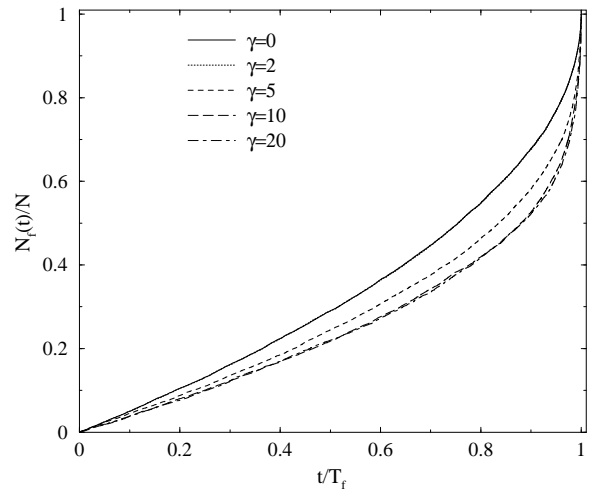,width=2.6in,angle=-90,clip=1}
\end{center}
\caption{Normalized number of broken fibers as a function of time
(also normalized to the lifetime of the bundle). Two groups of curves
can be clearly distinguished corresponding to the long range and short
range regimes. The results have been obtained for a system of $N=2500$
fibers and $\rho=2$.}
\label{fig3}
\end{figure}
Accordingly, the $\delta$s would be modified and there would be more
breakings for the {\em same} time interval. In other words, the
breakdown of the material occurs suddenly for the very localized
regime where about $50 \%$ of the fibers breaks in a time interval
of the order of $0.1 T_f$. In Fig.\ \ref{fig3} we have represented the
evolution with time of the number of broken fibers $N_f$ for different
values of $\gamma$. Two distinct groups of curves corresponding to the
extreme cases can be clearly seen. For intermediate values of the
effective range of interaction, the behavior is more like the case of
long range interaction and may correspond to other load sharing
schemes such as the hierarchical fiber bundle model \cite{r18,r25}.

\begin{figure}[t]
\begin{center}
\epsfig{file=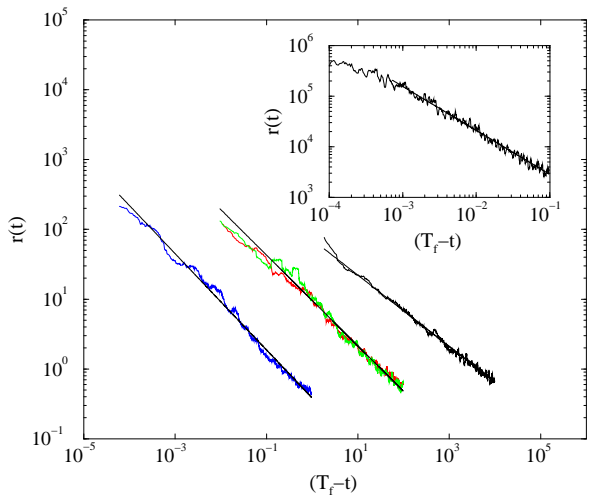,width=2.6in,angle=-90,clip=1}
\end{center}
\caption{Scaling of the breaking rate $r(t)$ (as defined in the text)
when approaching the time of material's breakdown. The values of
$\gamma$ are, from right to left, $0$, $4$, $5$, and $20$. With the
exception of $\gamma=20$, the curves have been shifted to the right for
the sake of clarity. The least square fit to the data gives for the
exponent the values $\zeta(\gamma=0)=0.5\pm 0.02$,
$\zeta(\gamma=4)=\zeta(\gamma=5)=0.58\pm 0.02$ and
$\zeta(\gamma=20)=0.66\pm 0.02$. The values of $N$ and $\rho$ are as
of Fig\ \ref{fig3}. The inset shows the same quantity for $\gamma=0$
and $\rho=10$. The scaling exponent is in this case $\zeta=0.86\pm
0.02$.}
\label{fig4}
\end{figure}

Consider that the breaking rate of the bundle is defined as
\begin{equation}
r(t)=\frac{dN_f(t)}{dt},
\label{eq7}
\end{equation}
with $N_f(0)=0$ and $N_f(T_f)=N$. Upon approaching the complete failure,
the breaking rate scales with the lifetime of the material as \cite{r26}
\begin{equation}
r(t)\sim (T_f-t)^{-\zeta},
\label{eq8}
\end{equation}
where the exponent $\zeta$ depends on both the range of interaction
and the Weibull index. However, we can again identify two limiting
groups of curves for the same value of $\rho$. Figure\ \ref{fig4}
shows the rate of fiber breaking for several load transfer ranges and
a Weibull index $\rho=2$. These results confirm the behavior observed
for the evolution of the number of broken fibers, namely that there is
a sharp increase in the failure rate when approaching the lifetime for
the case where short-range interaction dominates the damage
spreading. The fit to the curves gives $\zeta\approx 1/2$ for $\gamma
\leq \gamma_c$ and $\zeta\approx 2/3$ when $\gamma$ is in the range
where the effective interaction among the fibers can be assumed to be
very localized. Note that the curves for intermediate values of
$\gamma$ and for the GLS regime have been shifted for the sake of
clarity. The numerical results are not very smooth because of
fluctuations but the general trend of $r(t)$ confirms the validity of
relation\ (\ref{eq8}). As to the dependence of the above results on
the heterogeneity level we have observed that the less heterogeneous
the material is, the sharper the failure acceleration is in all cases.
The inset in Fig.\ \ref{fig4} shows the breaking rate for the case of
long range interaction and $\rho=10$. The higher value of
$\zeta\approx 0.86$ indicates that much of the fibers break in a very
small time interval close to the lifetime of the material. As the
range of interaction gets more localized, the exponent $\zeta$
approaches unity.

\section{Crack growth}
\label{scg}

A further characterization of what is going on in the fracture process
can be carried out by focusing on the properties of the clusters of
broken fibers. Specifically, we have monitored the growth of the
cracks inside the bundle. Cracks are defined as connected clusters or
regions of broken fibers. Here, we consider a coordination number
$q=8$, that is, each fiber has 8 neighbors. Similar results are
obtained if we take into account only nearest neighbors ($q=4$). At
the very initial stages, regardless of the range of interaction among
the fibers, the failure of fibers can be assumed to be random, that
is, the initial cracks are randomly nucleated inside the
material. This situation changes with time for different load transfer
schemes. By studying the growth of the largest crack area $C_m$ at
each time step, one could distinguish the different mechanisms leading
to the rupture of the material as the range of interaction varies. It
is worth noting that this is just a way that allows to discern between
different ranges of interaction and levels of heterogeneity. For
instance, one can consider instead the linear size of the largest
crack and study the fractal dimension of the crack distribution for
different $\rho$.

\begin{figure}[t]
\begin{center}
\epsfig{file=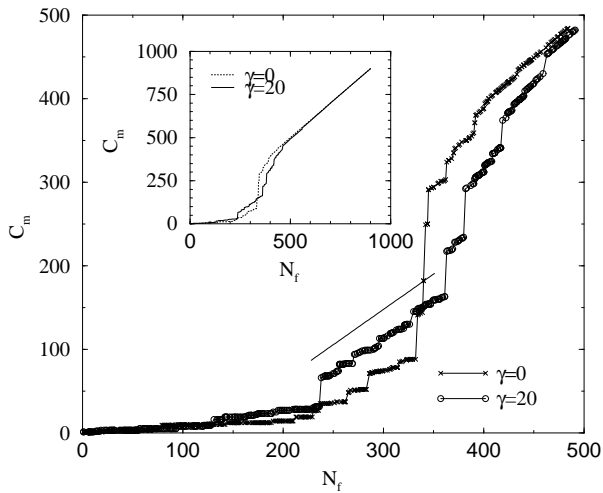,width=2.6in,angle=-90,clip=1}
\end{center}
\caption{Growth of the largest crack area $C_m$ with increasing number
of failed fibers $N_f$ for the two extreme load sharing rules here
illustrated for $\gamma=0$ (long range interaction) and $\gamma=20$
(short range interaction). The bundle consists of $N=900$ fibers and
$\rho=2$. The straight line is a fit to the form $C_m=a N_f$ with
$a=0.90 \pm 0.04$. The inset shows the evolution of $C_m$ up to the
macroscopic breakdown of the system for the same set of parameters.}
\label{fig5}
\end{figure}

Figure\ \ref{fig5} shows a microscopic aspect of the material
breakdown with time for the two limiting cases of load redistribution
and a system consisting of $N=900$ fibers ($\rho=2$).  Initially, it
can be seen that cracks are randomly nucleated in the material. At
later times the individual microcracks coalesce thereby causing a jump
in the largest crack area. For long range interaction, the nucleation
of cracks continues to be random because all the fibers carry the same
load and thus they break by chance. Therefore the largest broken
cluster does not grow linearly with the number of broken fibers. In
this case, there are isolated cracks inside the material that grow
essentially by coalescence when they meet each other. This is the
reason why sudden jumps in the area of the largest crack are observed
in the intermediate stages of the damage spreading. At the end of the
process, the material has accumulated many of these cracks giving rise
to the linear crack growth shown in the inset.

For localized range of interaction, the mechanism of damage spreading
is radically different. Again, at small times, the cracks are randomly
nucleated inside the material. However, as time goes on and more
fibers get broken, the load is redistributed to the fibers located
along the crack tips provoking the accumulation of stress in these
fibers and the appearance of regions where fibers bear a huge amount
of load. It is thus expected that the newly broken fibers add to
already existing cracks and that a dominant crack appears. From this
perspective, the largest crack area should scale linearly with the
number of broken fibers, {\em i.e.}, $C_m \approx N_f$. This is indeed
the case for $\gamma=20$ as represented in Fig.\ \ref{fig5}. The
straight line is a linear fit to a time window in which more than 150
fibers have been broken. The value $0.90$ of the slope confirms the
above picture. Note that in this case the number of coalescence events
is smaller than for the GLS regime and that after one of such events
the linear growth of the largest crack is recovered. We also note
that, up to the intermediate stages of the rupture process, the
largest crack for a given number of failed fibers is much higher in
the localized case than the global case indicating the formation of a
(few) dominant crack(s) in the later case. Additionally, at the end of
the process there are no differences between the two extreme load
transfer schemes since more than a half of the material is already
broken and is very unlikely to find a region where isolated cracks can
be formed and grow. Thus, at the final stage each additional breaking
event occurs at the crack tips of existing single dominant
cracks. Nevertheless, as stated in the previous section the rate of
fiber failures is quite different in both asymptotic regimes. We shall
note here that the same behavior as for the mean field regime is
observed for any value of $\gamma$ provided it is below the transition
point $\gamma_c=2$.

\begin{figure}[t]
\begin{center}
\epsfig{file=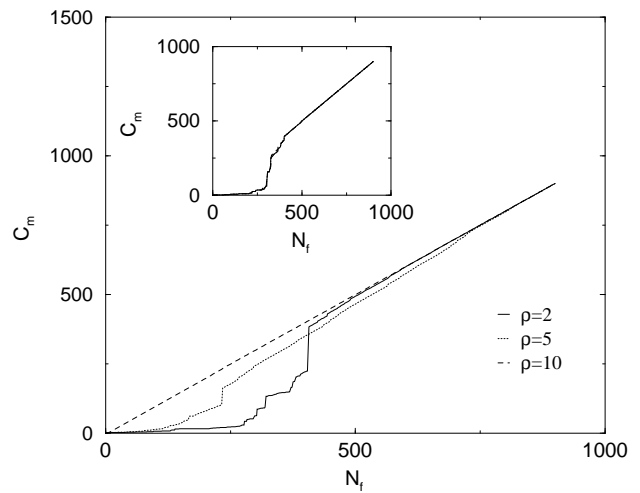,width=2.6in,angle=-90,clip=1}
\end{center}
\caption{Growth of the largest crack area $C_m$ with increasing number
of failed fibers $N_f$ for the local load sharing regime here
illustrated for $\gamma=10$ and several heterogeneity levels. The bundle
consists of $N=900$ fibers. The inset shows the evolution of $C_m$ for the
long range interaction regime ($\gamma=0$) and the same levels of
heterogeneity.}
\label{fig6}
\end{figure}

Figure\ \ref{fig6} further substantiates our previous arguments by
showing how the largest crack area varies as a function of the number
of broken fibers for several levels of heterogeneity. While for
$\gamma=0$ the picture is always the same (inset), for $\gamma=10$,
that is, in the localized regime, the time taken for cracks to become
dominant decreases with increasing $\rho$. Furthermore, when the
system is very homogeneous ($\rho=10$) and local interactions prevail,
a dominant crack which grows until the material collapses is formed
almost instantaneously confirming that the mechanism of rupture and
crack growth for homogeneous materials is radically different from that
of heterogeneous systems. Nevertheless, for the global load case, the
change in system's homogeneity does not alter the way dominant cracks
appear and grow (see the inset, where no changes, apart form
statistical fluctuations, can be observed). The reason why this
happens is given by the way the system gets broken. Equation\
(\ref{eq4}) tell us that the more stressed a fiber is, the more likely
it breaks. This always holds except for the global load sharing case,
where the fibers share the same amount of load and thus all of them
have the same probability to break. As the system is more homogeneous
(larger $\rho$), for local load sharing schemes, the probability of
breaking for the same load is higher so that the appearance of a
dominant crack is enhanced. Therefore, for long range interaction
there is no correlated crack growth while for short range regimes this
is precisely the dominant mechanism since the first stages of the
damage spreading.

\section{Conclusions}

We have extended the fiber bundle model with variable range of
interaction between fibers to the dynamic setting. As for the static
version, two very different regimes are identified as the exponent of
the stress transfer function varies. The lifetime of the material for
$\gamma < \gamma_c$ does not depend on both the system size and
$\gamma$ qualifying for a mean field behavior. On the contrary, for
the short range regime, the time to failure of the system
systematically decreases when increasing the size of the bundle. The
analysis performed also showed that the exact value of the crossover
point depends on the level of heterogeneity of the system. Besides, we
investigated how the material approaches its point of total breakdown
at both sides of the crossover point. The crossover from one regime to
the other also influences the behavior of the rate at which fibers
break, explicitly manifested in a power law divergence as $T_f$ is
approached, but with an exponent that depends on the range of
interaction. This result is relevant from a practical point of view
since for the localized regime the acceleration of the failure process
takes place at the very final stages of the rupture process. Although
fibers break by fatigue, one by one, they do break in very different
time intervals according to the range of interaction. In this sense, a
global load sharing regime is safer, since we get more warnings before
the material breakdown. On the other hand, the precursory activity
when the range of interaction gets localized is almost absent leading
to a sudden breakdown of the bundle in a very short time interval.

The numerical exploration of the damage spreading mechanisms under
different load transfer regimes further supported the results obtained
for the lifetime of the system. Regardless of the range of
interaction, the breaking of fibers is a completely random process at
the initial stages of the failure process. After some time, the
mechanism of failure propagation radically changes when the exponent
$\gamma$ varies. In the limiting case of global load sharing there is
no correlated crack growth in the system, whereas for the short range
regime the damage spreading is driven by a dominant crack, and thus, the
crack growth is strongly correlated with high stress concentration at
the fibers located along the perimeter of the dominant cluster of
broken fibers. 
\begin{figure}[t]
\begin{center}
\epsfig{file=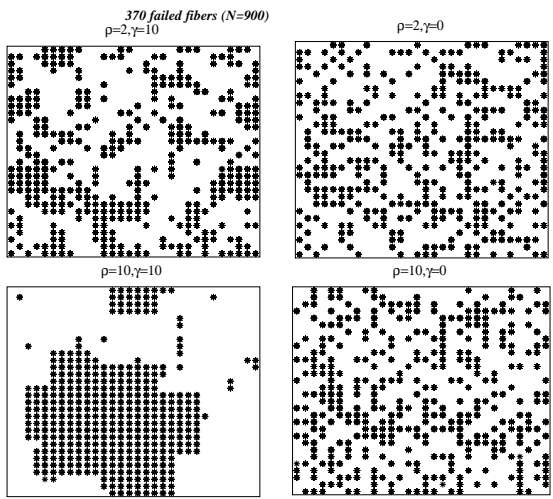,width=2.6in,angle=-90,clip=1}
\end{center}
\caption{Snapshots of the clusters when nearly $40\%$ of the
system is broken. The bundle consists of $900$ fibers and the
parameters of the model are shown above the figure. Note that
varying the heterogeneity level for $\gamma=0$ does not alter the
way in which nucleation and crack growth take place. For localized
regimes, as the system gets more homogeneous, the dominant crack
appears at early stages and drives the whole breakdown of the material.}
\label{fig7}
\end{figure}
These differences are clearly appreciated in Fig.\ \ref{fig7}, where
we represent snapshots of the lattice when nearly $40\%$ of the system
is broken for several values of $\gamma$ and $\rho$. For the long
range interaction limit the material's level of heterogeneity does not
influence the random nucleation and growth of cracks and there are no
clearly distinguishable dominant cracks. This continues to be so until
coalescence drives the further breaking of the material. On the
contrary, for localized regimes (left column in Fig.\ \ref{fig7}),
when the system gets more homogeneous the dominant crack appears at
early stages and damage spreading is strongly correlated resembling a
single crack growth mechanism typical of homogeneous materials. When
the bundle is heterogeneous, crack growth is still correlated, but in
this case we can identify more than one large and dominant
crack. Finally, our results suggest that actually there are only two
limiting cases relevant to experiments. The one in which mean field
assumptions apply could be of great importance since this will allow
the extension of known analytic results to ranges of interaction
beyond $\gamma=0$.

\section{Acknowledgments}

One of the authors (Y.\ M.\ ) would like to thank A.\ F.\ Pacheco and
J.\ B.\ G\'omez for valuable comments on this work. O.\ E.\ Y.\ thanks
the ICTP and UNESCO for their financial support and hospitality. Y.\
M.\ acknowledges financial support from the Ministerio de Educaci\'on
Cultura y Deportes (Spain) and of the Spanish DGICYT Project
BFM2002-01798. F.\ K.\ acknowledges financial support of the B\'olyai
J\'anos Fellowship of the Hungarian Academy of Sciences and of the
Research Contracts FKFP 0118/2001 and T037212. This work was partially
supported by the project SFB381, and by the NATO grant PST.CLG.977311.



\end{document}